\documentclass[conference,letterpaper]{IEEEtran}

\usepackage[utf8]{inputenc} 
\usepackage[T1]{fontenc}
\usepackage{url}
\usepackage{ifthen}
\usepackage{cite}
\usepackage{amsfonts}
\usepackage{indentfirst}
\usepackage{amsmath,amsthm}
\usepackage{amssymb}
\usepackage{color}
\usepackage{graphicx}
\usepackage[para]{threeparttable}
\usepackage{cases}
\usepackage{flushend}

\DeclareMathOperator{\Rnk}{Rank}
\DeclareMathOperator{\Norm}{Norm}

\usepackage{todonotes}

\usepackage[linesnumbered,algoruled,boxed,lined]{algorithm2e}

\newcommand{\etal}{\textit{et al.}}

\newcommand{\Fm}{\mathbb F_{q^m}}
\newcommand{\Fn}{\mathbb F_{q^n}}
\newcommand{\Fq}{\mathbb F_{q}}

\newtheorem{Theorem}{Theorem}

\newtheorem{Definition}{Definition}
\newtheorem{Proposition}{Proposition}

\newtheorem{Lemma}{Lemma}

\usepackage{threeparttable}
\usepackage{multirow}

\begin{document}
\title{On interpolation-based decoding of a class of maximum rank distance  codes} 


\author{%
  \IEEEauthorblockN{Wrya K. Kadir and Chunlei Li}
  \IEEEauthorblockA{Department of Informatics\\ University of Bergen, Norway\\
Email: \{wrya.kadir,chunlei.li\}@uib.no}
  \and
  \IEEEauthorblockN{Ferdinando Zullo}
  \IEEEauthorblockA{Dipartimento di Matematica e Fisica\\ Università degli Studi della Campania “Luigi Vanvitelli”, Italy\\
                    Email: ferdinando.zullo@unicampania.it}
}


\maketitle

\begin{abstract}
In this paper we present an interpolation-based decoding algorithm to decode a family of maximum rank distance codes proposed recently by Trombetti and Zhou. We employ the properties of the Dickson matrix associated with a linearized polynomial with a given rank and the modified Berlekamp-Massey algorithm in decoding. 
When the rank of the error vector attains the unique decoding radius, the problem is converted to solving a quadratic polynomial, which ensures that the proposed decoding algorithm has polynomial-time complexity.
\end{abstract}

\section{Introduction}

Rank metric codes were independently introduced by Delsarte \cite{Delsarte:1978aa}, Gabidulin \cite{Gabidulin1985} and Roth \cite{roth1991maximum}. Those rank metric codes that achieve Singleton-like bound are called \textit{maximum rank distance (MRD) codes}. The well known family of MRD codes are the \textit{Gabidulin codes}.
Later this family was generalized by Kshevetskiy and Gabidulin  \cite{kshevetskiy2005new} which is known as the \textit{generalized Gabidulin (GG)} codes. These codes are linear over $\Fn$. Sheekey \cite{Sheekey} introduced a large family of $\Fq$-linear MRD codes called \textit{twisted Gabidulin (TG) codes}, which were extended to \textit{generalized twisted Gabidulin (GTG) codes} by employing arbitrary automorphism \cite[Remark 9]{Sheekey},\cite{LTZ}. Later additive MRD codes were proposed by Otal and {\"O}zbudak \cite{Otal2017} and they are known as \textit{additive generalized twisted Gabidulin (AGTG) codes}. AGTG codes contain all the aforementioned MRD codes as subfamilies. There are also some other MRD codes that are not equivalent to the above codes, for instance the non-additive MRD codes by Otal and  {\"O}zbudak \cite{Otal:2018aa}, new MRD codes by Sheekey \cite{Sheekey2020newMRD}, \textit{Trombetti-Zhou (TZ) codes} \cite{TrombettiZhou2019}, etc. For more constructions of MRD codes, please refer to \cite{Sheey2019}.

MRD codes have gained much interest in the last decades due to their wide applications in storage system \cite{roth1991maximum}, network coding \cite{SilvaKschischangKoetter} and cryptography \cite{GPT}. Efficient decoding of MRD codes is critical for their applications. There are different decoding approaches for Gabidulin codes. Gabidulin \cite{Gabidulin1985} presented decoding based on a linearized equivalent of the Extended Euclidean Algorithm. The generalized Berlekamp-Massey algorithm was given by Richter and Plass in \cite{Richter}. Later Loidreau \cite{Loidreau:2006aa} proposed the Welch-Berlekamp like algorithm to decode Gabidulin codes. Nevertheless, the above algorithms can not be directly applied to the new MRD codes with twisted evaluation polynomials. Randrianarisoa and Rosenthal in \cite{rosenthal2017decoding} proposed a decoding method for a subfamily of TG codes. Randrianarisoa in \cite{Tovohery2018} gave an interpolation-based decoding algorithm for GTG codes. He reduced the decoding problem to finding zeros of projective equations. Kadir and Li in \cite{Kadir-li20} applied the interpolation approach to decoding AGTG codes and studied the final projective equations in greater depth. Li \cite{Li2019} used a similar idea in decoding the non-additive partition MRD codes in \cite{Otal:2018aa}.

In this paper we propose an interpolation-based decoding algorithm for TZ codes. We also compare the interpolation-based decoding algorithms for MRD codes when the rank of the error vector reaches the unique decoding radius, which shows that decoding TZ codes requires less operations than decoding GTG and AGTG codes as the problem can be reduced to solving a quadratic equation.

\section{Preliminaries} 

\begin{Definition}
Let $q$ be a power of prime $p$ and $\Fn$ be an extension of the finite field $\Fq$. A $q$-polynomial is a polynomial of the form $L(x)=a_0x+a_1x^q+\cdots+a_{k-1}x^{q^{k-1}}$ over $\Fn$. If $a_{k-1}\neq 0$, then we say that $L(x)$ has $q$-degree $k-1$. 
 The set of these polynomials is denoted by $\mathcal{L}_k(\Fn)$.
\end{Definition}
When $q$ is fixed or the context is clear, it is also customary to speak of a \textit{linearized polynomial} as it satisfies the linearity property: $L(c_1x+c_2y)=c_1L(x)+c_2L(y)$ for any $c_1,c_2 \in\Fq$ and any $x,y$ in an arbitrary extension of $\Fn$. Hence a linearized polynomial $L(x)\in \mathcal{L}_k(\Fn)$ defines an $\Fq$-linear transformation $L$ from $\Fn$ to itself.
The rank of a nonzero linearized polynomial $L(x)=\sum_{i=0}^{n}a_ix^{q^i}$ over $\Fn$ is given by $\mbox{Rank}(L)=n-\mbox{dim}_{\Fq}(\mbox{Ker}(L))$, where $\mbox{Ker}(L)$ is the kernel of $L(x)$.

\begin{Proposition}\label{prop-Dickson-tovo}
Let $L(x)=\sum_{i=0}^{n-1}a_ix^{q^i}$  over $\Fn$ be a linearized polynomial with rank $t$. Then its associated Dickson matrix 
\small
\begin{equation}\label{EqDicksonmatrix}
D=\begin{pmatrix}
a_{i-j({\,\rm mod }n)}^{q^i}
\end{pmatrix}_{n\times n}=\begin{pmatrix}
a_0& a_{n-1}^{q}& \cdots& a_1^{q^{n-1}}\\
a_1& a_0^{q}&\cdots& a_2^{q^{n-1}}\\
\vdots&\vdots&\ddots&\vdots\\
a_{n-1}& a_{n-2}^{q}&\cdots& a_{0}^{q^{n-1}}
\end{pmatrix},
\end{equation}
\normalsize
has rank $t$ over $\Fn$. Moreover, any $t\times t$ submatrix formed by $t$ consecutive rows and $t$ consecutive columns in $D$ is non-singular. 
\end{Proposition} 
The first part of Prop. \ref{prop-Dickson-tovo} is given in \cite{WU201379}, whereas the second part can be found in \cite{Tovohery2018} and \cite{csajbok2020scalar}. 

\section{Maximum rank distance (MRD) codes}

The rank of a vector $a=(a_1,\ldots,a_n)$ in $\Fm^n$, denoted as $\mbox{Rank}(a)$, is the number of its  linearly independent components, that is the dimension of 
the vector space spanned by $a_i$'s over $\Fq$. The rank distance between two vectors $a,b\in \Fm^n$ is defined as $d_R(a,b)=\mbox{Rank}(a-b)$.

\begin{Definition}
A subset $\mathcal{C}\subseteq \Fm^n$ with respect to the rank distance is called a rank metric code. When $\mathcal{C}$ contains at least two elements, the minimum rank distance of $\mathcal{C}$ is given by $d(\mathcal{C})=\displaystyle\min_{\substack{A,B\in \mathcal{C},~A\neq B}}\{{{d_R}}(A,B)\}$. Furthermore, it is called a \textit{maximum rank distance (MRD) code} if it attains the Singleton-like bound $|\mathcal{C}|\leq q^{\min \{m(n-d+1),n(m-d+1)\}}$. 
\end{Definition}
The most famous MRD codes are Gabidulin codes \cite{Gabidulin1985} which were further generalized in \cite{roth1996tensor,kshevetskiy2005new}. The generalized Gabidulin (GG) codes $\mathcal{GG}_{n,k}$ with length $n\leq m$ and dimension $k$ over $\Fm$ is defined by the evaluation of 
\begin{equation}\label{eq-GG codes}
\bigg\{\sum_{i=0}^{k-1}f_ix^{q^{si}}\; |f_i\in \Fm\bigg\},
\end{equation}
where $(s,{m})=1$, on linearly independent points $\alpha_0,\alpha_1,\ldots,\alpha_{n-1}$ in $\Fm$.
The choice of $\alpha_i$'s does not affect the rank property and it is customary to exhibit Gabidulin codes and its generalized families without the evaluation points as in  \eqref{eq-GG codes}. For consistency with the parameters of MRD codes in \cite{Sheekey,TrombettiZhou2019,Otal2017}, through what follows we always assume $n=m$.

For a linearized polynomial $L(x)=\sum_{i=0}^{k}l_ix^{q^i}$ over $\Fn$, it is clear that $\mbox{Rank}(L)\geq n-k$ if $l_{k}\neq 0$. Gow and Quinlan in \cite[Theorem 10]{GOW20091778} (see also \cite{Sheekey}) characterize a necessary condition for $L(x)$ to have rank $n-k$ as below, see \cite{CSAJBOK2019109lin-poly,MCGUIRE201968} for other necessary conditions. 
\begin{Lemma}\label{Sheekey-lemma} \cite{GOW20091778}
		Suppose a linearized polynomial $L(x)=l_0x+l_1x^{q}+\cdots+l_kx^{q^{k}}$, $l_k\neq 0$, in $\mathcal{L}_n({\mathbb{F}_{q^n}})$ has $q^k$ roots in $\mathbb{F}_{q^n}$. Then 
		$\Norm_{q^n/q}(l_k)= (-1)^{nk}\Norm_{q^n/q}(l_0),$
		where $\Norm_{q^n/q}(x)=x^{1+q+\cdots + q^{n-1}}$ is the norm function from $\mathbb{F}_{q^n}$ to $\mathbb{F}_q$.
\end{Lemma}
According to Lemma \ref{Sheekey-lemma}, a linearized polynomial $L(x)$ of $q$-degree $k-1$ has rank at least $n-k+1$ if the condition in Lemma \ref{Sheekey-lemma} is not met. 
Sheekey \cite{Sheekey} applied Lemma \ref{Sheekey-lemma} and constructed a new family of MRD codes, known as \textit{twisted Gabidulin (TG) codes}, and the generalized TG codes are investigated in \cite{LTZ}.
Later Otal and {\"O}zbudak \cite{Otal2017} further generalized this family by manipulating some terms of linearized polynomials and constructed the \textit{additive generalized twisted Gabidulin (AGTG) codes} which contains all the aforementioned MRD codes as subfamilies. 

Below we recall from \cite{TrombettiZhou2019} the \textit{Trombetti-Zhou (TZ) code}, which has been proved to be inequivalent to subfamilies of AGTG codes, further generalized twisted Gabidulin codes \cite{sven-sheekey-newMRD}, Sheekey's new MRD codes \cite{Sheekey2020newMRD} and those with minimum distance equals to $n-1$, such as \cite{CsMPZ,CSAJBOK18-newMRD}. We are going to propose an interpolation-based decoding algorithm for TZ codes in the next section.  

	\begin{Proposition}
	
	 \cite{TrombettiZhou2019}
		Let $n,k,s\in \mathbb{Z}^+$ satisfying $(s,2n)=1$ and let $\gamma\in \mathbb{F}_{q^{2n}}$ satisfy that $\mbox{Norm}_{q^{2n}/q}(\gamma)$ is a non-square element in $\mathbb{F}_{q}$.  Then the set 
	\small	
		
		\begin{equation*}\label{EqTZ}
		\mathcal{D}_{k,s}(\gamma)= \left\{ ax+\sum_{i=1}^{k-1}f_ix^{q^{si}} + \gamma b x^{q^{sk}} \;|\; f_i \in \mathbb{F}_{q^{2n}}, a,b\in \mathbb{F}_{q^n} \right\}
		\end{equation*}
		\normalsize
is an $\mathbb{F}_{q^n}$-linear MRD code of size $q^{2nk}$ and minimum rank distance $2n-k+1$.
	\end{Proposition}
	 The first and the last coefficients of the above polynomial are chosen independently from the base field $\mathbb{F}_{q^n}$. If $q$ is even, all the elements of $\Fq$ are square elements, so TZ codes exist  only when the characteristic of $\Fq$ is odd.

	\section{Encoding and decoding of TZ codes}

For the rest of this paper, we will denote $[i]:=q^{si}$ for $i=0, \ldots, 2n-1$ , where $(s,2n)=1$, for simplicity.

\vspace{-6pt}
\subsection{Encoding}\label{Subsec3.1}

	For a TZ MRD code with evaluation points $\alpha_0,\alpha_1, \ldots, \alpha_{2n-1}$ that are linearly independent over $\mathbb{F}_{q}$, the encoding of
	a message $f=(f_0,\ldots, f_{k-1})$ is the evaluation of the following linearized polynomial at points $\alpha_0,\alpha_1, \ldots, \alpha_{2n-1}$:
	\begin{equation}\label{eq-TZ-linearized-form}
	    f(x)=ax+\sum_{i=1}^{k-1}f_ix^{[i]}+\gamma b x^{[k]},
	\end{equation} where $(a, b)\in \mathbb{F}_{q^n}
	\times \mathbb{F}_{q^n}$ corresponds to $f_0$ via an $\mathbb{F}_{q^n}$-basis of $\mathbb{F}_{q^{2n}}$. 
	Let 
	$\tilde{f}=(a,f_1,\ldots, f_{k-1}, \gamma b, 0, \ldots, 0)$ be a vector of length $2n$ over $\mathbb{F}_{q^{2n}}$ and $	M=
	\begin{pmatrix}
		\alpha_i^{[j]}
	\end{pmatrix}_{2n\times 2n}$
	be the $2n\times 2n$ \textit{Moore matrix} generated by $\alpha_i$'s, where $1\leq i, j\leq 2n-1$.

	Then the encoding of TZ codes can be expressed as 
	\small
	\begin{equation}\label{EqTZ-Encoding}
	(a,f_1,\ldots, f_{k-1}, \gamma b)\mapsto c=(f(\alpha_0), \ldots, f(\alpha_{2n-1}))=\tilde{f}M^T,
	\end{equation}
	\normalsize
	where $M^T$ is the transpose of matrix $M$.	Here it is worth noting that  in encoding process, one actually only needs to 	calculate the multiplication of the $(k+1)$-tuple $(a,f_1,\ldots, f_{k-1}, \gamma b)$ and the first $k+1$ rows of $M$.
		Here we express it as in \eqref{EqTZ-Encoding} for being consistent with the decoding procedure.

	\vspace{-6pt}
	\subsection{Decoding}\label{Subsec3.2}

	For a received word $r=c+e$ with an error $e$ added to the codeword $c$ during transmission, 
	when the error $e$ has rank $t\leq \lfloor \frac{2n-k}{2}\rfloor$,
	the unique decoding task is to recover the unique codeword $c$ such that $d_R(c,r)\leq \lfloor \frac{2n-k}{2}\rfloor$.

		\smallskip

	\smallskip
	
	Suppose $g(x)=\sum_{i=0}^{2n-1}g_ix^{[i]}$ is an error interpolation polynomial such that
	\begin{equation}\label{EqInterpolation}
	g(\alpha_i)=e_i=r_i-c_i, \quad i=0, \ldots, 2n-1.
	\end{equation}
	It is clear that the error vector  $e$ is uniquely determined by the polynomial $g(x)$ and denote $\tilde{g}=(g_0,\ldots, g_{2n-1})$. From
	\eqref{EqTZ-Encoding} and \eqref{EqInterpolation} it follows that
	$$
	r = c+e = (\tilde{f}+\tilde{g}) M^T.
	$$
	This is equivalent to
	\begin{align*}
	  r \cdot (M^T)^{-1}=& (a,\, f_1, \,\ldots, f_{k-1}, \gamma b, 0, \ldots, 0) + \\
	  & (g_0,g_1, \ldots, g_{k-1}, g_k, g_{k+1}, \ldots, g_{2n-1}).
		\end{align*}

	Letting $\beta=(\beta_0, \ldots, \beta_{2n-1})=r \cdot (M^T)^{-1}$, we obtain 
	\begin{equation} \label{EqInterpolation-3}
	(g_{k+1}, \ldots, g_{2n-1}) = (\beta_{k+1}, \ldots, \beta_{2n-1}) 
	\end{equation}
	and
\begin{equation*}
\begin{cases}
g_0+a=\beta_0\\
g_k+\gamma b=\beta_k
\end{cases}\rightarrow \begin{cases}
g_0-\beta_0=-a\\
\gamma^{-1}(g_k-\beta_k)= -b.
\end{cases}
\end{equation*}
With $a, b\in \mathbb{F}_{q^n}$, one obtains 
\begin{equation}\label{eq-g0 gk1}
\begin{cases}
(g_0-\beta_0)^{[n]}=g_0-\beta_0\\
(\gamma^{-1}(g_k-\beta_k))^{[n]}=\gamma^{-1}(g_k-\beta_k).
\end{cases}
\end{equation}
\normalsize
which yields two linearized equations
    \begin{numcases}{}
    g_0^{[n]}-g_0-\theta_1=0 \label{eq-g0 gk2-1},\\
    g_k^{[n]}-\gamma^{[n]-1}g_k-\theta_2=0 \label{eq-g0 gk2-2},
    \end{numcases}
 where $\theta_1=\beta_0^{[n]}-\beta_0,\;\theta_2 = \beta_k^{[n]}  -\gamma^{[n]-1}\beta_k$.

	\medskip

	Therefore, the task of correcting error $e$ is equivalent to 
	reconstructing $g(x)$ from the available information characterized in \eqref{EqInterpolation-3}, \eqref{eq-g0 gk2-1} and \eqref{eq-g0 gk2-2}. This reconstruction process heavily depends on the 
	property of the associated  Dickson matrix of $g(x)$ and 
	will be discussed in Subsection \ref{SubSec3.3}.

	\subsection{Reconstructing the interpolation polynomial $g(x)$}\label{SubSec3.3}

	\smallskip

	The Dickson matrix associated with $g(x)$ can be given by 
	\begin{equation}\label{Eq-DM-G-Simplified}
	G=\begin{pmatrix}g^{[j]}_{i-j~({\rm mod~}2n)} 
	\end{pmatrix}_{2n\times 2n}
	= \left(G_0 \,\, G_1 \,\, \ldots \,\, G_{2n-1}\right),
	\end{equation} 
	where the indices $i, \,j$ run through $\{0, 1, \ldots, 2n-1\}$ and 
	$G_{j}$ is the $j$-th column of $G$.

Since $\gcd(2n, s)=1$, Proposition \ref{prop-Dickson-tovo} can be easily adapted for the Dickson matrix $G$ in \eqref{Eq-DM-G-Simplified}.
Hence $G$ has rank $t$ and any $t\times t$ matrix formed by $t$ successive rows and columns in $G$ is nonsingular. 
Then $G_0$ can be expressed as a linear combination 
of $G_1, \ldots, G_t$, namely,
$
G_{0} = \lambda_1 G_1 +\lambda_2 G_2 + \cdots + \lambda_{t} G_{t},
$
where $\lambda_1,\ldots, \lambda_{t}$ are elements in $\mathbb{F}_{q^{2n}}$.
This yields the following recursive equations
\begin{equation}\label{Eq-Gsub}
g_{i} = \lambda_1 g^{[1]}_{i-1} + \lambda_2 g^{[2]}_{i-2} + \cdots + \lambda_t g^{[t]}_{i-t}, \quad 0\leq i < 2n,
\end{equation} where the subscripts in $g_i$'s are taken modulo $2n$.
Recall that the elements $g_{k+1},\ldots, g_{2n-1}$ are known from \eqref{EqInterpolation-3}. 
Hence we obtain the following linear equations with known coefficients and  variables $\lambda_1, \ldots, \lambda_{t}$:
\begin{equation}\label{Eq-Gsub-4}
g_{i} = \lambda_1 g^{[1]}_{i-1} + \lambda_2 g^{[2]}_{i-2} + \cdots + \lambda_t g^{[t]}_{i-t}, \,\, k+t+1\leq  i <2n.
\end{equation} 
	The above recurrence gives a generalized version of $q$-linearized shift register as described in \cite{Sidorenk}, 
	where $(\lambda_1, \ldots, \lambda_{t})$ is the connection vector of the shift register.
    It is  the \textit{key equation} for the decoding algorithm in this paper, by which we shall
	reconstruct $g(x)$ in two major steps: 
	\begin{itemize}
		\item[]\noindent\textbf{Step 1.} derive  $\lambda_1, \ldots, \lambda_{t}$ from 
		\eqref{EqInterpolation-3}-\eqref{eq-g0 gk2-2}, and \eqref{Eq-Gsub-4};
		\item[]\noindent\textbf{Step 2.} use $\lambda_1, \ldots, \lambda_{t}$ to compute $g_{k}, \ldots, g_0$ from \eqref{Eq-Gsub}.
	\end{itemize}
	Step 1 is the critical and challenging step in the decoding process, and Step 2 is simply a recursive process that can be done in linear time in $\mathbb{F}_{q^{2n}}$.
	The following discussion shows how the procedure of  Step 1 works. 
	
	As discussed in the beginning of this section, for an error vector with $\Rnk(e)=t \leq \lfloor \frac{2n-k}{2}\rfloor$, i.e., $2t+k\leq 2n$, we can divide the discussion into two cases.
	
	\noindent\textit{Case 1:} $2t+k<2n$. In this case, \eqref{Eq-Gsub-4} contains 
	$2n-k-t-1\geq t$ affine equations 
	in variables $\lambda_1, \ldots, \lambda_{t}$, which has rank $t$. Hence the variables $\lambda_1, \ldots, \lambda_{t}$ can be uniquely determined. In this case, the code can be seen as a sub-code of an $\mathcal{GG}_{2n,k+1}$ code and any Gabidulin codes decoding algorithm is applicable.	Here we assume the code has high code rate, for which the Berlekamp-Massey algorithm is more efficient. In addition it is consistent with the notation used in Case 2. 
	Although the recurrence equation \eqref{Eq-Gsub-4} is a generalized version of the ones in \cite{Richter,Sidorenk}, the modified Berlekamp-Massey algorithm can be applied here to recover the 
	coefficients $\lambda_1, \ldots, \lambda_{t}$.

	\smallskip
	
	\noindent\textit{Case 2:} $2t+k=2n$. In this case \eqref{Eq-Gsub-4} 
	gives $2n-k-t-1=t-1$ independent affine equations in variables $\lambda_1, \ldots, \lambda_{t}$.
	For such an under-determined system of linear equations, we will have a set of 
	solutions $(\lambda_1, \ldots, \lambda_{t})$
	that has dimension $1$ over $\mathbb{F}_{q^{2n}}$. 
	Namely, the solutions will be  of the form 
	$$
	\lambda+\omega \lambda'
	=(\lambda_1+\omega\lambda'_1, \ldots, \lambda_{t}+\omega \lambda'_{t}),
	$$ where $\lambda, \lambda'$ are fixed elements in $\mathbb{F}_{q^{2n}}^t$ and $\omega$ runs through $\mathbb{F}_{q^{2n}}$.
	As shown in \cite[Th. 10]{Sidorenk}, the solution can be derived from the modified BM algorithm with a free variable $\omega$.
	Next we will show how the element $\omega$ is determined by other  information in \eqref{EqInterpolation-3}, \eqref{eq-g0 gk2-1} and \eqref{eq-g0 gk2-2}.

	\smallskip
	
	Observe that in \eqref{Eq-Gsub}, by taking $i=0$ and $i=k+t$ and substituting the solution $\lambda+\omega\lambda'$, one gets the following two equations
	$$
	\begin{array}{rcllll}
	g_{0}&=& (\lambda_1+\omega \lambda'_1, \ldots, \lambda_{t}+\omega \lambda'_{t}) \cdot (g_{2n-1}^{[1]}, \ldots, g_{2n-t}^{[t]})^T,
	\\
	g_{k+t}&=& (\lambda_1+\omega \lambda'_1, \ldots, \lambda_{t}+\omega \lambda'_{t}) \cdot (g_{k+t-1}^{[1]}, \ldots, g_{k}^{[t]})^T,
	\end{array}
	$$ 
	where $g_0,g_k$ and $\omega$ are the only unknowns. 

	Re-arranging the equations gives
	\begin{equation}\label{eq20}
	g_0  =  c_0 + c_1 \omega ,
	\end{equation}
	and
	\begin{equation} \label{eq21}
	g_{k+t} =  c_2+c_3 \omega + (\lambda_{t}+\lambda'_{t}\omega)g_{k}^{[t]}, 
	\end{equation}
	where $c_0, c_1, c_2, c_3$ are derived from $\lambda$, $\lambda'$ and the known coefficients $g_{i}$'s.
 	Furthermore, from \eqref{eq-g0 gk2-1} and \eqref{eq-g0 gk2-2} we have  $g_0^{[n]}-g_0+\theta_1=0$ and $g_k^{[n]}-\gamma^{[n]-1}g_k+\theta_2=0$. Substituting \eqref{eq20} in \eqref{eq-g0 gk2-1} gives 
 	\begin{equation}\label{eq22}
 	    c_1 \omega^{[n]} +\beta_1 \omega +\beta_2=0.
 	\end{equation}
 	If $\lambda_t+\lambda_t'\omega= 0$ then we have the solution $\omega=-\lambda_t/\lambda_t'$. This solution can be further checked in \eqref{eq21} by $g_{k+1}, c_2$ and $c_3$. Otherwise, one can raise both sides of \eqref{eq21} to the $[2n-t]$-th power and obtain

 	\begin{equation}\label{eq24}
 	    g_k=\dfrac{a_1+a_2\omega^{[2n-t]}}{a_3+a_4\omega^{[2n-t]}}.
 	\end{equation}
 	Replacing this value in \eqref{eq-g0 gk2-2}, raising it to the ${[t]}$-th power and rearranging the terms implies
 	\begin{equation}\label{eq27}
 	\zeta_1\omega^{[n]+1}+\zeta_2\omega^{[n]}+\zeta_3\omega+\zeta_4=0,
 	\end{equation}
 	where $\zeta_1 =(a_2^{[n]}a_4+\theta_2a_4^{[n+t]})^{[t]}$.
 	Furthermore, by \eqref{eq22} and \eqref{eq27} we have the following quadratic equation over $\mathbb{F}_{q^{2n}}$
 	\begin{equation}\label{eq28}
 	\zeta_1x^2+\zeta_5x+\zeta_6=0.
 	\end{equation}

When $\zeta_1=0$ and $\zeta_2\neq 0$, the unknown $\omega$ can be uniquely determined.
  When $\zeta_1\neq 0$, the above quadratic equation 
 can be reduced to 
 \begin{equation}\label{eq29}
 x^2+rx+s=0,
 \end{equation} 
 where $r=\zeta_5/\zeta_1$ and $s=\zeta_6/\zeta_1$.
  
  Since the characteristic of $\mathbb{F}_q$ is odd,  Equation \eqref{eq29} can be solved explicitly as follows: 
\begin{itemize}
    \item[a)] if ${r^2-4s}$ is a quadratic residue in  $\mathbb{F}_{q^{2n}}$, then it has two solutions $x=\frac{-r\pm\sqrt{r^2-4s}}{2}$;
    \item[b)] if $r^2=4s$, then it has a single solution $x=-r/2$;
    \item[c)] it has no solution in      $\mathbb{F}_{q^{2n}}$  otherwise.
\end{itemize}

	\smallskip

	Since the error $e$ with rank $t=\frac{2n-k}{2}=\frac{d-1}{2}$ can be uniquely decoded, our quadratic equation
	should have roots $w$ in $\mathbb{F}_{q^{2n}}$ that lead to solutions $\lambda+\omega \lambda'$ in \eqref{Eq-Gsub-4} and $(g_0, g_k)$ in \eqref{eq20}. With the coefficients $\lambda_1, \ldots, \lambda_{t}$ in Step 1
	and the initial state $g_{2n-1}, \ldots, g_{2n-t}$, 
	one can recursively compute
	$g_0, \ldots, g_{k-1}$  according to \eqref{Eq-Gsub} in Step 2.
	Note that even if the equation \eqref{eq28} has two different solutions, 
	they don't necessarily lead to correct coefficients of the error interpolation polynomial.
	In fact, by the expression of Dickson matrix of $g(x)$, the correct 
	$g(x)$ should have the sequence $(g_{2n-1}, \ldots, g_{2n-t}, \ldots )$ generated from \eqref{Eq-Gsub}  has period $2n$.
	In other words, if the output sequence has period $2n$, we know that
	the corresponding polynomial $g(x)=\sum_{i=0}^{2n-1}g_ix^{[i]}$ is the desired error interpolation polynomial. For self-completeness, the decoding process of TZ codes is summarized in Algorithm \ref{Alg1}.
	\begin{algorithm}[http]\label{Alg1}
		\small
		\SetAlgoLined
		\KwIn{A received word $r$ with $t\leq \lfloor \frac{2n-k}{2}\rfloor$ errors and linearly independent evaluation points $\alpha_1, \ldots, \alpha_{2n}$}
		\KwOut{The correct codeword $c\in \mathbb{F}_{q^{2n}}^n$ or  ``Decoding Failure"}
		Calculate $\beta(x)=\sum_{i=0}^{2n-1}\beta_i x^{[i]}$ such that $\beta(\alpha_i) = r_i$ for $i=1, \ldots, 2n$\;
		Apply modified BM algorithm to $(g_{k+1}, \ldots, g_{2n-1})=(\gamma_{k+1}, \ldots, \gamma_{2n-1})$ and output $L$, $\Lambda^{(2n-k-1)}(x)$, $B^{(2n-k-1)}(x)$\;
		\If{$L=(2n-k)/2$}
		{   
			Denote  $\Delta=\omega+\sum_{i=1}^{L}\Lambda_i^{(2n-k-1)}g_{2n-1-i}^{q^{si}}$ with $\omega\in \mathbb{F}_{q^{2n}}$ \;
			Express the coefficients of the polynomial
			$$
			\Lambda^{(2n-k)}(x)=\Lambda^{(2n-k-1)}(x) -\frac{1}{\Delta}  x^{q^s}\circ B^{(2n-k-1)}(x),
			$$Derive the vector $\lambda+\lambda'\omega$ by negating the coefficients of $\Lambda^{(2n-k)}(x)$;
			
			\eIf{$\lambda_t+\lambda_t'\omega=0$}{
    $\omega=-\lambda_t/\lambda_t'$\;
   }{
   Derive the polynomial $P(x)=\zeta_1x^2+\zeta_5x+\zeta_6$ as in \eqref{eq28}\;
 	\eIf{$\zeta_1\neq 0$}{Solve $P(x)=0$ by Cases a)-c) after \eqref{eq29}\;}{The zero of $P(x)$ is $x=\zeta_6/\zeta_5$\;} }
		
			Set $(\lambda_1, \ldots, \lambda_{t})=\lambda+\omega \lambda'$ with $\omega$ as the zero of $P(x)$\;
			Calculate $g_0, g_k$ from \eqref{eq20} and \eqref{eq21}\;
		}
		\For{each $i$ in $\{0, \ldots, k\}$}{
			Calculate 
			$g_i=\lambda_1g_{i-1}^{[1]} +\cdots +\lambda_{t}g_{i-t}^{[t]}$, where the subscripts of $g_j$'s are taken modulo $2n$\;
		}
		\eIf{The sequence $g_0,\ldots, g_{2n-1}$
 		derived from $\lambda_1, \ldots, \lambda_t$
		has period $2n$}
		{Return the codeword $c=(c_0, \ldots, c_{2n-1})$ with $c_i=r_i-g(\alpha_i)$}
		{Return ``Decoding Failure"}
		\caption{Interpolation decoding of TZ codes}
		\normalsize
	\end{algorithm}

\subsection{Complexity Analysis}
As summarized in Algorithm \ref{Alg1}, we have two major steps to construct the error interpolation polynomial $g(x)$. The first step is to use the modified BM algorithm for obtaining the coefficients $\lambda_1,\ldots,\lambda_t$. Calculating the interpolation polynomial at points $(\alpha_i,r_i)$ has complexity in the order of $\mathcal{O}(n^3)$, but according to \cite{Gao1993}, if $\alpha_0,\ldots,\alpha_{2n-1}$ is taken as a self-dual normal basis, $M$ is orthogonal, which means $M^T=M^{-1}$ and computation of $(M^T)^{-1}$ is no longer required. So the complexity of computing polynomial $\beta$ is reduced to $\mathcal{O}(n^2)$ over $\mathbb{F}_{q^{2n}}$. The second major component of the first step is the modified BM algorithm which is known to have complexity in the order of $\mathcal{O}(n^2)$ over $\mathbb{F}_{q^{2n}}$. The second step is to deal with the case $t=\lfloor(2n-k)/2\rfloor$ by investigating the solutions of the equation \eqref{eq28}. This step involves checking whether $(r^2-4s)$ is a quadratic residue or not. In order to check whether an element $a\in \mathbb{F}_{q^{2n}}$ is square or not, one calculates $a^{\frac{q^{2n}-1}{2}}=a^{\frac{q-1}{2}\cdot(q^{2n-1}+\cdots+q+1)}=b^{q^{2n-1}+\cdots+q+1}$ which has complexity $\mathcal{O}(n)$ over $\mathbb{F}_{q^{2n}}$, or directly check its exponent if in implementation an element in $\mathbb{F}_{q^{2n}}$ is represented in exponential form. 
As a result, the complexity of our decoding method is in the order of $\mathcal{O}(n^2)$ over $\mathbb{F}_{q^{2n}}$.

Therefore, the previous two sections imply the following result.
\begin{Theorem}
	Consider the evaluation code obtained from $\mathcal{D}_{k,s}(\gamma)$ over an $\mathbb{F}_q$-basis of $\mathbb{F}_{q^{2n}}$. Every received word can be uniquely decoded up to rank $t\leq \frac{2n-k}{2}$ errors in polynomial time.
	\end{Theorem}

\section{Comparing the known decoding algorithms}
Known decoding algorithms for Gabidulin codes can be generally classified in two different approaches: syndrome decoding  as in \cite{Gabidulin1985,GPT,roth1991maximum,Richter} and interpolation-based decoding as in \cite{Loidreau:2006aa,Tovohery2018,Kadir-li20,Li2019,Li2018}. When the rank of the error vector reaches the maximal unique decoding radius, syndrome decoding approach works only for $\Fn$-linear MRD codes.
Since Sheekey \cite{Sheekey} introduced TG codes, which is not always $\Fn$-linear, a new (non syndrome) decoding algorithm for rank metric codes has been required for the extreme case when $t=\lfloor \frac{n-k}{2}\rfloor$. When the rank of the error is not the maximal unique decoding radius, i.e., $t<\lfloor \frac{n-k}{2}\rfloor$,  the syndrome decoding algorithms are still applicable. 
 Loidreau \cite{Loidreau:2006aa} proposed the first interpolation-based decoding approach for MRD codes and considered the analogue of Welch-Berlekamp algorithm, which was  originally used to decode Reed-Solomon codes.
Later Randrianarisoa \cite{Tovohery2018} employed Berlekamp-Massey algorithm as the main seed and introduced a decoding algorithm for GTG codes. Later Kadir and Li \cite{Li2018,Kadir-li20} used the same idea to decode AGTG codes. In the rest of this section, we compare the existing interpolation-based decoding algorithms for MRD codes when $t=\lfloor\frac{n-k}{2}\rfloor$.

 The goal of the WB algorithm is to find two linearized polynomials $V$ and $N$ with $q$-degrees less than or equal to $t$ and less than $k+t$,  respectively, which satisfy the system of equations $V(r_i)-N(\alpha_i)=0$ where $i=1,\ldots,n$. The system is a linear system consists of $n$ equations and $n+1$ unknowns. This is equivalent to interpolating two pairs of linearized polynomials $(V_0,N_0)$ and $(V_1,N_1)$. After an initialization step, the polynomials are interpolated via a loop with indices ranging from $k$  to $n-1$. If one manages to bound the $q$-degree of the polynomials as $\mbox{deg}_q(V_j)\leq t$ and $\mbox{deg}_q(N_j)\leq k+t-1$ for $j=0$ or $1$, it is done. The complexity of the WB algorithm is in the order of $\mathcal{O}(n^2)$ over $\Fn$.

  The decoding algorithms in \cite{Tovohery2018} and \cite{Kadir-li20}  interpolated the polynomial $f(x)+g(x)$ where $f(x)$ and $g(x)$ correspond to message vector $c$ and error vector $e$, respectively. The decoding problem is reduced to the problem of solving  an under-determined system of linear equations with $t-1$ equations and $t$ unknowns. This approach benefits from the properties of Dickson matrix associated with $g(x)$, known coefficients of $g(x)$ and the relation between $f_0$ and $f_k$ which enable us to convert the system of equations to a single projective polynomials of the form $P(x)=x^{q^v+1}+u_1x+u_2=0$ for GTG and AGTG codes. The zeros of this polynomial were discussed in \cite{Kadir-li20} when $(v,n)=1$. Very recently Kim \etal in  \cite{kim-sihem2021complete} provide the complete solution of $P(x)=0$ over $\Fn$ for any power prime $q$ and any integers $n$ and $v$. Note that the relation between the coefficients of the first and the last terms of $f(x)$ in the decoding algorithm for TZ codes provides more useful information than the corresponding equations for GTG and AGTG codes. It turns out that we only need to deal with a quadratic polynomial instead of a projective  polynomial.   This makes the decoding algorithm for TZ codes faster than decoding GTG and AGTG codes.



\section{Conclusion}
In this paper we proposed an interpolation-based decoding algorithm for Trombetti-Zhou MRD codes. We have shown that the decoding algorithm has polynomial time complexity as low as $\mathcal{O}(n^2)$ over $\mathbb{F}_{q^{2n}}$. It involves Berlekamp-Massey algorithm similar to the decoding approaches in \cite{Tovohery2018,Kadir-li20} but end up with a quadratic polynomial, rather than a projective polynomial, which requires less operations ($\mathcal{O}(n)$)  to compute the zeros.

\bibliographystyle{IEEEtran}
\bibliography{RankMetricCodes}

\end{document}